\documentclass[twocolumn,aps,showpacs]{revtex4}
\usepackage{graphicx}
\usepackage{amsmath}
\usepackage{amsfonts}
\usepackage{amssymb}

\begin{document}

\title{\textbf{Complex network study of Brazilian soccer players }}
\author{Roberto N. Onody}
\email{onody@if.sc.usp.br}
\author{Paulo A. de Castro}
\email{pac@if.sc.usp.br}
\affiliation{Departamento de F\'{\i}sica e Inform\'atica,
Instituto de F\'{\i}sica de S\~ao Carlos, \\
Universidade de S\~ao Paulo, C.P.369, 13560-970 S\~ao Carlos-SP,
Brazil}

\begin{abstract}
Although being a very popular sport in many countries, soccer has not received much
attention from the scientific community. In this paper, we study soccer from a complex
network point of view. First, we consider a bipartite network with two kinds
of vertices or nodes: the soccer players and the clubs. Real data were gathered from
the $32$ editions of the Brazilian soccer championship, in a total of $13,411$ soccer players
and $127$ clubs. We find a lot of interesting and perhaps unsuspected results.
The probability that a Brazilian soccer player has worked at $N$ clubs
or played $M$ games shows an exponential decay while the probability that he has scored
$G$ goals is power law. Now, if two soccer players who have worked at the same club at the same time are
connected by an edge, then a new type of network arises (composed exclusively by soccer
players nodes). Our analysis shows that for this network the degree distribution
decays exponentially. We determine the exact values of the clustering coefficient,
the assortativity coefficient and the average shortest path length and compare them with
those of the Erd\"os-R\'enyi and configuration model. The time evolution of these quantities
are calculated and the corresponding results discussed.


\end{abstract}
\maketitle

In the past few years there has been a growing interest in the study of complex networks.
The boom has two reasons - the existence of interesting applications in several biological,
sociological, technological and communications systems and the availability of a large
amount of real data \cite{new0,dm,domuwa,ws}.

Social networks are composed by people interacting with some pattern of contacts like
friendship, business or sexual partners. One of the most popular works in this area was
carried out by Milgram \cite{mil} who first arrived to the concept of the "six degrees of
separation" and small-world. Biological networks are those built by nature in its
indefatigable fight to turn life possible: the genetic regulatory network for the
expression of a gene \cite{lee}, blood vessels \cite{wbe}, food webs \cite{cgar} and
metabolic pathways \cite{rsmo}. Technological or communications networks are those
constructed by man in its indefatigable fight to turn life good: electric power grid
\cite{ws,asbs}, airline routes \cite{asbs}, railways \cite{sen}, internet \cite{fff} and
the World Wide Web \cite{baj}.

A network is a set of vertices or nodes provided with some rule to connect them by edges. The degree
of a vertex is defined as being equal to the number of edges connected to that vertex.
In order to characterize a network, six important quantities or properties
can be calculated \cite{new0,dm}: the degree distribution, the clustering coefficient, the assortativity
coefficient, the average shortest path length, the betweenness and the robustness to a failure or attack.
The first four quantities appear in this work and their meanings are explained below.

In this report, we study a very peculiar network: the Brazilian soccer network. Using the information
at our disposal \cite{pla}, a bipartite network is constructed with two types
of vertices: one composed by $127$ clubs (teams) and the other formed by $13,411$ soccer players. They
correspond to the total number of clubs and soccer players that have sometime participated of the Brazilian
soccer championship during the period $1971-2002$ \cite{cbf}.
Whenever a soccer player has been employed by a certain
club, we connect them by an edge.

\begin{figure}[htbp!]
\begin{center}
\includegraphics[width=7cm]{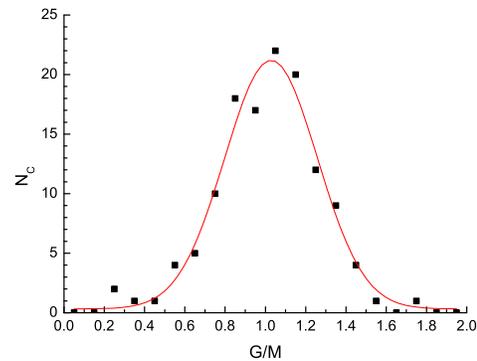}
\end{center}
\caption{Histogram of the number of clubs against the number of goals scored
by match. Bins of size $0.1$ were used.
The full line corresponds to the fitted Gaussian curve. The average number of
goals by match is equal to $1.00$.}%
\label{Figure1}%
\end{figure}

Figure 1 shows the number of clubs $N_{C}$ versus $G/M$, which stands
for goals by match and it is equal to the total number of goals
($G$) scored by a club, divided by the number of matches ($M$) disputed by that club.
Clearly, the data are well fitted by a
Gaussian curve centered at $G/M \simeq 1.03$. The Brazilian club
with the best index is S\~ao Caetano ($G/M = 1.73$) and the worst is
Colatina with $G/M = 0.22$.

Figure 2 plots the degree distributions for each kind of vertex
of the bipartite network: players and clubs.
The player probability $P(N)$ exhibits an exponential decay with the player degree $N$.
Naturally, $N$ corresponds to the number of clubs in which a player has ever worked.
We find the average $\overline{N} = 1.37$.
The most nomad player
is Dad\'a Maravilha \cite{nick} with $N = 11$.
The inset shows the club probability $P(S)$ as a function of the club degree $S$.
Regrettably, its form cannot be inferred may be because the
small number of involved clubs.

\begin{figure}[htbp!]
\begin{center}
\includegraphics[width=7.5cm]{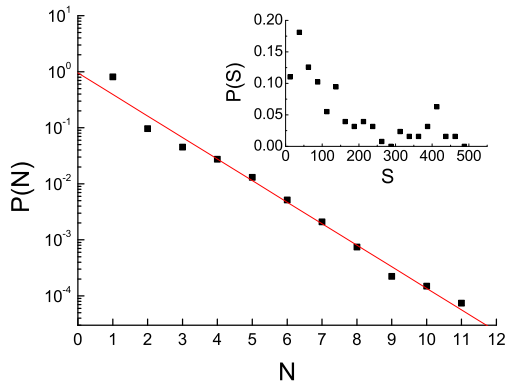}
\end{center}
\caption{Probability $P(N)$ that a player has
worked for $N$ clubs. The full line corresponds to the
fitted curve $P(N) \sim 10^{-0.38 N}$.
So, it is $190$
times more probable to find someone who has played for only two
clubs than for eight clubs. The inset is the degree distribution
$P(S)$ for the clubs.}%
\label{Figure2}%
\end{figure}

\begin{figure}[htbp!]
\begin{center}
\includegraphics[width=8cm,height=7.5cm]{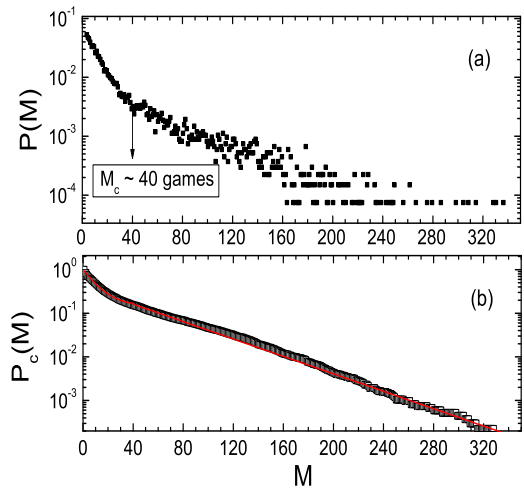}
\end{center}
\caption{(a) Game probability $P(M)$ versus the number $M$
of disputed matches. (b) Cumulative distribution
$P_{c}(M)$ built from $P(M)$. Fittings appear as
the full lines.}%
\label{Figure3}%
\end{figure}

A very amazing result comes out when we determine the probability
$P(M)$ that a soccer player has played a total of $M$ games
(disregarding by what club). There is an elbow (see Figure 3a) or a critical value
at $M_{c} = 40$ for the semi-log plot of $P(M)$.
As there is a lot of scatter, we also determine the corresponding
cumulative distribution $P_{c}(M)$ \cite{new0}.
The latter distribution is very well fitted by two different exponentials:
$P_{c}(M) = 0.150 + 0.857 \; 10^{-0.042 M}$	 for $M < 40$ and
$P_{c}(M) = 0.410 \; 10^{-0.010 M}$ for $M > 40$,
as it is shown in the Figure 3b.  This implies that the original distribution
$P(M)$ has also two exponential regimes with the {\it same} exponents \cite{new0}.
The existence of the threshold $M_{c}$ probably indicates that, after a
player has found some fame or notoriety, it is
easier to him to keep playing soccer. Surpassing this value is like the player
has gained some kind of "stability" in his job.

\begin{figure}[htbp!]
\begin{center}
\includegraphics[width=8cm,height=8.5cm]{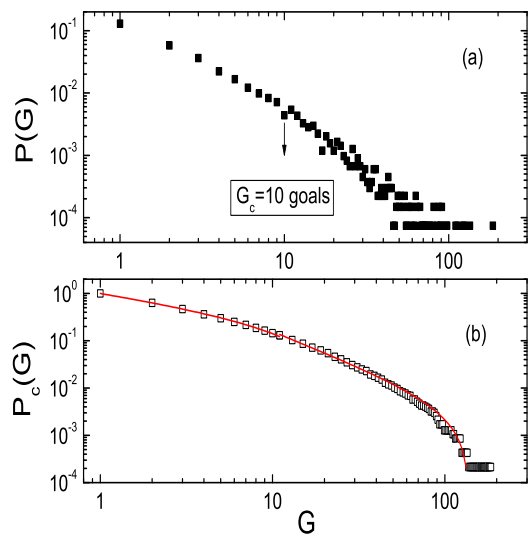}
\end{center}
\caption{(a) Goals probability $P(G)$ that a player has
scored $G$ goals.
Choosing randomly, a Brazilian soccer player has ten times less chance
to have scored $36$ goals than $13$ goals. The player with the
highest score is Roberto Dinamite with $G = 186$.
(b) The corresponding cumulative
probability distribution $P_{c}(G)$.}%
\label{Figure4}%
\end{figure}

As the goals are the quintessence of the soccer, we determine the
goals probability $P(G)$ that a player has scored $G$
goals in the Brazilian championships. The result is shown in the Figure 4a.
Here again we find an intriguing threshold at $G_{c} = 10$ separating
regions with apparently two distinct power law exponents. Such kind of
behavior has already been found in the context of scientific collaborations network
\cite{new1}. To verify if
this threshold really exist (since the tail is, once more, very scattered) we calculate
the corresponding cumulative distribution $P_{c}(G)$, plotted in the Figure 4b.
The curve we get resembles that one found in the network of collaborations
in mathematics (see fig. 3.2(a) of ref. 1) and it may correspond to a truncated power-law
or possibly two separate power-law regimes.
We have tried to fit $ P_{c}(G)$ with truncated power laws.
However, our best result was obtained using two power laws with different exponents:
$P_{c}(G) = -0.259 + 1.256 \; G ^{-0.500}$ for $G < 10$
and $P_{c}(G) = -0.004 + 4.454 \; G^{-1.440}$ if $G > 10$. Notice that the existence of
additive constants in the power laws spoils the expected straight line characteristic
in the log-log plot. Moreover, as the cumulative distribution does not preserve
power law exponents, it follows that $P(G) \sim G^{- 1.5}$ and $P(G) \sim G^{- 2.44}$ for $G < 10$
and $G >10$, respectively.
We conjecture that the origin
of this threshold can be simply explained by the structure,
position or distribution of the players in the soccer field.
Circa of two thirds of the eleven players form in the defense or
in the middle field. Players in these positions usually score less
than those of the attack.

From the bipartite network (of players and teams), one can construct an unipartite network composed
exclusively by the soccer players. If two players were at the {\it same team} at
the {\it same time}, then they will be connected by an edge. Let us call the resulting network as the
Brazilian soccer player (BSP) network.
With this merging, we get a time growing network that reflects acquaintances and possible social relationships
between the players. Similar merging have already been done for the bipartite networks: actor-film \cite{ws,asbs},
director-firm \cite{davis} and scientist-paper \cite{new1}.

In the year 2002, the BSP network had $13,411$ vertices (the soccer players)
and $315,566$ edges. The degree probability distribution $P(k)$
can be easily calculated and we obtain an average degree $\overline{k} = 47.1$.
The result is plotted in the Figure 5.

\begin{figure}[htbp!]
\begin{center}
\includegraphics[width=7cm]{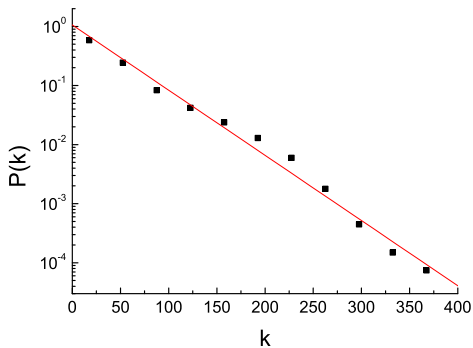}
\end{center}
\caption{Degree distribution of the BSP network.
The fitting curve (full line) has the exponential form
$P(k) \sim 10^{-0.011k}$.}%
\label{Figure5}%
\end{figure}

Many others quantities of the BSP network
can be precisely evaluated. One can measure, for example,
what is the probability $C_{i}$ that the first neighbors of a vertex $i$
are also connected. The average of this
quantity over the whole network gives the clustering
coefficient $C$, a relevant parameter in social networks. We
find $C = 0.79$, which means that the BSP is a highly clustered network.
At this point, it is very interesting to compare the BSP network results with
those of random graphs as the Erd\"os - R\'enyi (ER) model \cite{er} and the configuration model
\cite{new0,new2}. We simulated a ER network (with the same size as the BSP) in which
the vertices are connected with a probability equal to $0.00351$, which gives, approximately, the same number
of edges for both networks. Still keeping the network size, we also simulated the configuration model
using the fitting curve of the Figure 5 as the given degree distribution. We see from Table 1,
that both ER and the configuration model have a small clustering coefficient as it would
be expected for random networks.

\begin{center}
\begin{table}[htbp!]
\begin{tabular}{|c|r|r|r|} \hline
\multicolumn{1}{|c|}{} & \multicolumn{1}{|c|}{BSP}
& \multicolumn{1}{|c|}{Erd\"os-R\'enyi} & \multicolumn{1}{|c|}{Configuration}
 \\ \hline
$v$ & $13,411$ & $13,411$ & $13,411$ \\ \hline
$e$ & $315,566$ & $315,443$ & $345,294$ \\ \hline
$\overline{k}$ & $47.1$ & $47.1$ & $51.5$ \\ \hline
$C$ & $0.790$ & $0.004$ & $0.008$ \\ \hline
$A$ & $0.12$ & $0.00$ & $0.46$ \\ \hline
$D$ & $3.29$ & $2.84$ & $2.85$ \\ \hline
\end{tabular}
\caption{In the first column, $v$ is the number of vertices, $e$ is the number of edges,
$\overline{k}$ is the mean connectivity, $C$ is the clustering coefficient, $A$ is the assortativity
coefficient and $D$ is the average shortest path length. }
\end{table}
\end{center}

The assortativity coefficient
$A$ \cite{new0} measures the tendency of a network to connect
vertices with the same or different degrees. If $A > 0$ ($A <
0$) the network is said assortative (disassortative)  and no
assortative when $A = 0$. To determine $A$, we first need to
calculate the joint probability distribution $e_{jk}$, which is the
probability that a randomly chosen edge has vertices with degree $j$
and $k$ at either end. Thus,

\begin{equation}
A = \frac{1}{\sigma^{2}_{q}} \sum_{jk} j k (e_{jk} - q_{j}q_{k}),
\end{equation}
where $ q_{k} = \sum_{j} e_{jk}$  and $\sigma_{q}^{2} = \sum_{k}
k^{2} q_{k} - ( \sum_{k} k q_{k} ) ^{2}$  . The possible values of
$A$ lie in the interval	 $-1 \leq A \leq 1$. For the BSP network,
we find $A = 0.12$ so it is an assortative network. This value
coincides with that of the Mathematics coauthorship \cite{new3}
and it is smaller than that of the configuration model ($A = 0.46
$). The explanation is very simple: although the vertices of the
configuration model are in fact randomly connected, the
given degree distribution constraint generates very strong correlations.
This can be measured by the nearest-neighbors average connectivity of a vertex with
degree $k$, $K_{nn}(k)$ \cite{sat}, which is plotted in Figure 6.

\begin{figure}[htbp!]
\begin{center}
\includegraphics[width=7cm]{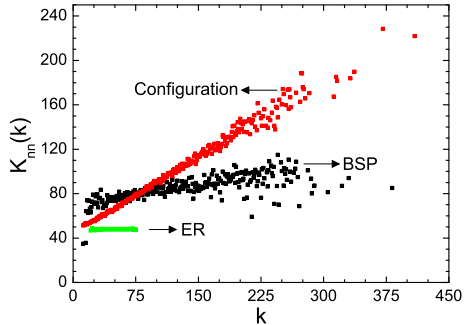}
\end{center}
\caption{Nearest-neighbors average connectivity for the ER, BSP
and Configuration model.}%
\label{Figure6}%
\end{figure}

Finally, we also determine the average shortest path length $D$ between a given vertex and all the others
vertices of the network. Taking the average of this quantity for all BSP network vertices, we get $D = 3.29$.
In analogy with social networks, we can say that there is {\it 3.29 degrees of separation}
between the Brazilian soccer players or, in other words, the BSP network is a small-world.

We can also study the time evolution of the BSP network. We have verified that
this network is broken in many clusters in 1971 and 1972, after that there is only one component.
In Table II, we observe an increasing mean
connectivity $\overline{k}$. We can think of two reasonings for that: the player's professional life is
turning longer and/or the players transfer rate between teams is growing up. On the other hand, the clustering
coefficient is a time decreasing function. Also in this case, there may be two possible explanations: the
players transfer rate between national teams and the exodus of the best Brazilian players to foreigner teams (which
has increased, particularly, in the last decades).

\begin{center}
\begin{table}[htbp!]
\begin{tabular}{|c|r|r|r|r|r|r|} \hline
\multicolumn{1}{|c|}{} & \multicolumn{1}{|c|}{1975}
& \multicolumn{1}{|c|}{1980} & \multicolumn{1}{|c|}{1985}
& \multicolumn{1}{|c|}{1990} & \multicolumn{1}{|c|}{1995}
& \multicolumn{1}{|c|}{2002}
 \\ \hline
$v$ & $2,490$ & $6,420$ & $8,797$ & $10,329$ & $11,629$ & $13,411$ \\ \hline
$e$ & $48,916$ & $128,424$ & $181,293$ & $219,968$ & $254,371$ & $315,566$	\\ \hline
$\overline{k}$ & $39.3$ & $40.0$ & $41.2$ & $42.6$ & $43.7$ & $47.1$  \\ \hline
$C$ & $0.84$ & $0.83$ & $0.82$ & $0.81$ & $0.80$ & $0.79$  \\ \hline
$A$ & $0.02$ & $0.06$ & $0.06$ & $0.07$ & $0.08$ & $0.12$  \\ \hline
$D$ & $3.17$ & $3.35$ & $3.39$ & $3.27$ & $3.28$ & $3.29$  \\ \hline
\end{tabular}
\caption{Temporal evolution of some quantities of the BSP network. The meanings of
the first column are the same as those of Table I.}
\end{table}
\end{center}

Naturally, this kind of movement diminishes the
cliques probabilities. From the Table II, we also see that the BSP network is becoming more assortative with time.
This seems to indicate the existence of a growing segregationist pattern, where the players transfer occurs,
preferentially, between teams of the same size. Finally, the average shortest path length values may suggest
that it is size independent but, most probably, this conclusion is misled by the presence of only
some few generations of players in the growing BSP network.

We hope that the work presented here may stimulate further researches in this subject. Some
opened questions are, for instance, whether the results obtained for the Brazilian soccer
held for different countries or, perhaps, for different sports.

We acknowledge financial support by CNPq (Conselho Nacional de Desenvolvimento Cient\'{\i}fico e Tecnol\'ogico)
and FAPESP (Funda\c c\~ao de Amparo a Pesquisa do Estado de S\~ao Paulo). We are very grateful to Marli Schapitz
de Castro and Henrique Jota de Paula Freire for help us to manipulate the database.


\begin{thebibliography}{99}

\bibitem{new0} M. E. J. Newman, SIAM Review \textbf{45}, 167
(2003).

\bibitem{dm} S. N. Dorogovtsev and J. F. F. Mendes in {\it Evolution of
Networks}, Oxford University Press (2003).

\bibitem{domuwa} P. S. Dodds, R. Muhamad and D. J. Watts, Science \textbf{301}, 827 (2003).

\bibitem{ws} D. J. Watts and S. H. Strogatz, Nature (London)
\textbf{393}, 440 (1998).

\bibitem{mil} S. Milgram, Psycology Today \textbf{1}, 60 (1967).

\bibitem{lee} T. I. Lee et al., Science \textbf{298}, 799 (2002).

\bibitem{wbe} G. B.West, J. H. Brown and B. J. Enquist, Nature \textbf{400}, 664 (1999).

\bibitem{cgar} J. Camacho, R. Guimer\'a and L. A. N. Amaral, Phys.
Rev. Lett. \textbf{88}, 228102 (2002).

\bibitem{rsmo} E. Ravasz, A. L. Somera, D. A. Mongru, Z. N. Oltvai and A.-L Barab\'asi,
Science \textbf{297}, 1551 (2002).

\bibitem{asbs} L. A. N. Amaral, A. Scala, M. Barthe\'el\'emy and
H. E. Stanley, Proc. Natl. Acad. Sci. USA \textbf{97}, 11149 (2000).

\bibitem{sen} P. Sen et al., Phys. Rev. E \textbf{67}, 036106 (2003).

\bibitem{fff} M. Faloutsos, P. Faloutsos and C. Faloutsos,
Computer Communications Review \textbf{29}, 251 (1999).

\bibitem{baj} A.-L. Barab\'asi, R. Albert and H. Jeong, Physica A
\textbf{272}, 173 (1999).

\bibitem{pla} Data were gathered from the CD-ROM released by the magazine Placar
(Editora Abril, Brazil, 2003).

\bibitem{cbf} The first official Brazilian soccer championship was disputed in $1971$ and
since then it is managed by the CBF (Confedera\c c\~ao Brasileira de Futebol).

\bibitem{nick} We preferred to use the nicknames so that the fans can easily recognize
their idols. Few people knows that Pel\'e's name is \'Edson Arantes do Nascimento.

\bibitem{new1} M. E. J. Newman, Phys. Rev. E \textbf{64}, 016131
(2001).

\bibitem{davis} G. F. Davis, M. Yoo and W. E. Baker, Strategic Organization \textbf{3},
301 (2003).

\bibitem{er} P. Er\"os and A. R\'enyi, Publ. Math. \textbf{6}, 290
(1959).

\bibitem{new2} M. E. J. Newman, S. H. Strogatz and D. J. Watts, Phys. Rev. E \textbf{64},
026118 (2001).

\bibitem{new3} M. E. J. Newman, Phys. Rev. Lett. \textbf{89},
208701 (2002).

\bibitem{sat} R. Pastor-Satorras, A. V\'asquez and A. Vespignani, Phys. Rev. Lett. \textbf{87},
258701 (2001).



\end{thebibliography}
\end{document}